\documentclass[12pt]{article}
\usepackage{epsfig,graphicx}

\title{Magnetic susceptibility at zero and nonzero chemical potential  in QCD and QED}
\author{V.~D.~Orlovsky and  Yu.~A.~Simonov,\\ Institute of Theoretical and Experimental Physics\\
Bolshaya Cheremushkinskaya 25, Moscow 117218, Russia}
\date{}

\newcommand{\be}{\begin{equation}}
\newcommand{\ee}{\end{equation}}

\def\la{\mathrel{\mathpalette\fun <}}
\def\ga{\mathrel{\mathpalette\fun >}}
\def\fun#1#2{\lower3.6pt\vbox{\baselineskip0pt\lineskip.9pt
\ialign{$\mathsurround=0pt#1\hfil ##\hfil$\crcr#2\crcr\sim\crcr}}}

\newcommand{{\SD}}{\rm SD}

\newcommand{{\Mc}}{\mathcal{M}}

\newcommand{\lan}{\langle}

\begin{document}

\maketitle

\begin{abstract}
\noindent

 Magnetic susceptibility of the quark matter in QCD is calculated in a closed
 form for an arbitrary chemical potential $\mu$. For small $\mu$, $\mu\ll T$,
 $\sqrt{eB} \ll T$, a strong dependence on temperature $T$ is found due to
 Polyakov line factors. In the opposite case of small $T$, $\sqrt{eB} \ga T$,
 the oscillations as functions of $eB$ occur, characteristic of the  de
 Haas-van Alphen effect. Results are compared with available lattice data.

\end{abstract}
\section{Introduction}

The important role of magnetic  fields (m.f) in nature has recently become a
topic of a vivid  interest. Strong m.f. are expected in
cosmology \cite{1} and  in astrophysics (magnetars) \cite{2}, very large m.f.
can occur  in heavy-ion collisions \cite{3}, where a temperature transition to
the quark-gluon matter is expected. For a modern review of these topics see
\cite{4}.

For a theory the m.f. effects play an additional role  of a crucial test, which
proves or disproves the assumed intrinsic  dynamics, or shows its boundaries.

Of special importance strong m.f. are in QCD, since both hadrons   and the
quark matter are possible parts  of neutron stars and m.f. can occur in heavy
ion  collisions.

Recently a new approach was suggested  to  treat QCD and QED in m.f., based on
the relativistic Hamiltonians, deduced from the  QCD path integral
\cite{5,6,7}.

A striking result of this approach is the strong reduction of the hadron masses
due to m.f. in mesons \cite{8,9,10}, and in baryons \cite{11}. For the neutron
the mass is twice as small for $eB = 0.2$ GeV$^2$. These results for mesons are
supported by lattice data \cite{12,13,14}.

Another feature of m.f. is the lowering of the  temperature  $T_c$ of the
transition from the hadronic to the quark-gluon matter, which was found in the
same  path integral approach \cite{15,16} and supported by the lattice data
\cite{17}. It is a purpose  of the present paper to develop the theory further
and to find in a simple closed form   the magnetic susceptibility (m.s.) of the
quark matter for an arbitrary chemical potential $\mu$.  Recently this type of
analysis was done for the zero $\mu$ \cite{18} and the numerical results for
m.s. have been compared to lattice data \cite{19,19*,19**,20}, showing a good
agreement.

The nonzero $\mu$ case is interesting from several points of view. First of all
it covers the regions of $\mu, T, eB$ which are present also in the case of the
electron gas, and where the effects of the Pauli paramagnetism \cite{21} and
Landau diamagnetism \cite{22} occur, moreover there also the de Haas-van Alphen
phenomenon is possible, see the regular course \cite{23} for a general
discussion. As we shall see, in the case of the quark gas a simple modification
occurs in all these effects, and in addition another, and possibly more
important  region, $\mu\ll T$ exists, where our method allows to obtain simple
general results.

The paper is organized as follows. In the next section general expressions for
the thermodynamical potentials in m.f. are derived, in section 3 the expression
for the m.s. $\hat \chi$ is deduced. Section 4 is devoted to the m.s. at
nonzero  $\mu$, while in the section 5 the classical Pauli, Landau and de
Haas-van Alphen effects are demonstrated for the quark matter.

In section 6 the main results are summarized and perspectives are given.

\section{A general theory of the fermion gas in magnetic field}

We start with the case of the electron gas in m.f., where the
thermodynamical potential $\Omega (V,T,\mu)$ (or rather $\Omega
(T) - \Omega(0) \equiv -P$) , which we use in what follows, can be
written as \cite{23}

\be P_e (B, \mu, T) = \sum_{n_\bot ,\sigma} \frac{eB T}{2\pi} X(\mu)
,~~ X(\mu)=\int\frac{dp_z}{2\pi} \ln \left(1+ \exp \left(
\frac{\mu- E^\sigma_{n_{\bot}}(B)}{T}\right)\right), \label{1}\ee
where \be E^\sigma_{n_\bot} (B) = \sqrt{ p^2_z+ (2n_\bot +1 -
\sigma) eB + m^2_e}, ~~ \sigma = \pm 1.\label{2}\ee

Note, that (\ref{2}) is the relativistic generalization of the standard
expression \cite{23} in the theory of the electron (or electron-positron) gas
in m.f. at nonzero temperature. It  was a subject of an intensive study during
the last 50 years, see e.g. \cite{24,25,26,27,28}. The QED relativistic
thermodynamical potential in m.f. at finite $T$ and density was obtained in
\cite{24}, and using the generalized Fock-Schwinger method for $\mu\neq 0,
T\neq 0$ in \cite{25}. In the case of the zero temperature and nonzero $\mu, B$
the useful form of the effective action was obtained in \cite{26}, and finally
the full expression for nonzero,  $\mu, T, B$ was presented in \cite{27}.
Simple forms of effective Lagrangians for $T=0$ and oscillations as function of
m.f. are obtained in \cite{28}. For further developments and discussions and
limiting cases see also \cite{29}. These results have been exploited and
augmented by the study of the quark-antiquark gas also in magnetic field \cite{
31, 32,33}.

In the latter case one can write for a given sort of quarks and antiquarks
similarly to (\ref{1}), if one neglects the effect of the vacuum QCD fields on
quarks 
\be 
P_q (B, \mu, T) = \sum \frac{N_c e_q BT}{2\pi} (X_q (\mu) +
X_q (-\mu)),\label{3}
\ee 
and $X_q(\mu)$ has the same form as in (\ref{1}),
(\ref{2}) with $e=e_q\equiv |e_q|,$ and $m_e \to m_q$.

However, the vacuum QCD fields, which are responsible for confinement at $T<
T_c$ \cite{34}, also affect the quark gas. The theory of both confined and
deconfined matter was suggested in \cite{35} and finally formulated, basing on
the path integral formalism and the Field Correlator Method (FCM) in \cite{
36,37,38}, for a review see \cite{40}.

In this formalism, neglecting the $q\bar q$ weakly bound states around $T_c$
(the ``Single Line Approximation'' SLA \cite{36}) one arrives at the simple
modification of the expression (\ref{3}), where one should replace in
$X(\mu)$ the chemical potential $\mu$ as follows \be \exp \frac{\mu}{T} \to
\exp \frac{\mu_q}{T} L(T),\label{4}\ee where $L(T)= \exp \left( -
\frac{V_1(\infty, T)}{2T}\right)$ is the average value of the fundamental
Polyakov line, which was studied analytically in \cite{38,40} and numerically
on the lattice in \cite{41}.

As a result of  integration over $dp_z$ in $X(\mu)$ one  arrives at the
expression \cite{15}, containing a sum over Matsubara numbers \be P_q (B, \mu,
T) = \frac{N_c e_q BT}{\pi^2} \sum_{n_\bot, \sigma} \sum^\infty_{n=1}
\frac{(-)^{n+1}}{n} L^n \frac{e^{\frac{n\mu}{T}}+ e^{-\frac{ \mu n}{T}}}{T}
\varepsilon^\sigma_{n_\bot} K_1 \left( \frac{n
\varepsilon^\sigma_{n_\bot}}{T}\right)\label{5}\ee with $K_1 (z)$ -- the
modified Bessel function and \be \varepsilon^\sigma_{n_\bot} = \sqrt{ e_q B (2
n_\bot + 1 -\sigma) + m^2_q}.\label{6}\ee Another form of (\ref{5}) was
obtained in \cite{15} by direct summing $\sum\limits_{n_\bot, \sigma} X_q (\mu)$ in
(\ref{3}), which gives the integral expressions

 \be  P_q (B, \mu, T) = \frac{N_c e_q B}{2\pi^2} (\psi (\mu) + \psi (-\mu)
), ~~ \psi (\mu) = \phi (\mu) + \frac23 \frac{\lambda (\mu)}{e_q B} - \frac{e_q
B \tau (\mu)}{24}, \label{7}\ee where $\phi(\mu)$, $\lambda (\mu)$ and
$\tau(\mu)$ are integrals over momenta  given in (35),(36),(37).

In what follows we shall be mostly using the form (\ref{5}), which was summed
up over $n_\bot, \sigma$ for $\mu=0$ in \cite{15}
$$ P_q (B,\mu,T) =\frac{N_c e_q BT}{\pi^2}   \sum^\infty_{n=1}
\frac{(-)^{n+1}}{n} L^n \{ m_q K_1 \left( \frac{nm_q}{T}\right)+$$ \be +
\frac{2T}{n}
 \frac{e_q B + m^2_q}{e_q B} K_2
 \left( \frac{n}{T} \sqrt{e_q B + m^2_q}\right) - \frac {ne_q B}{12T} K_0 \left( \frac{n}{T} \sqrt{m^2_q + e_q B}\right)
 \}.\label{8}\ee

 It is easy to see, that the case of $\mu>0$ obtains by the formal replacement
 \be L^n \to L^n \textrm{ch} \left( \frac{ \mu n}{T} \right) = \frac{ L^n_\mu +
 L^n_{-\mu}}{2}, ~~ L_\mu \equiv e^{\frac{\mu}{T}} L.\label{9}\ee

 The form (\ref{8}) or its nonzero $\mu$  equivalent (\ref{9}) have a nice
 property of yielding correct limiting values for  1) $e_q B\to 0$, 2) $e_qB\to
 \infty$, 3) $T\gg m_q, e_qB$.

 In the first case only the second term inside  curly brackets in (\ref{8})
 contributes and one has
 \be P_q (0, \mu,T) =\frac{N_c 2T^2m^2_q}{\pi^2} \sum^\infty_{n=1}
 \frac{(-)^{n^{n+1}}}{n^2} \frac{L^n_\mu+ L^n_{-\mu}}{2} K_2 \left(
 \frac{nm_q}{T}\right).\label{10}\ee

 In the second case only the first  term inside   curly brackets survives and
 we   obtain
 \be P_q (B, \mu, T ) |_{B\to \infty} = \frac{ N_c e_q B T m_q}{\pi^2}
 \sum^\infty_{n=1} \frac{(-)^{n+1}}{n} \frac{ L^n_\mu+ L^n_{-\mu}}{2} K_1 \left(
 \frac{nm_q}{T}\right).\label{11}\ee

 At large $T$, $T\gg \sqrt{m^2_q+ e_q B}$, the leading term in (\ref{8}) is again the second in the curly
 brackets and one has

\be P_q (B, \mu, T\to \infty) = \frac{ 4N_c  T^4}{\pi^2}
 \sum^\infty_{n=1} \frac{(-)^{n+1}}{n^4} \frac{ L^n_\mu+
 L^n_{-\mu}}{2},\label{12}\ee which yields for $\mu=0, L=1$ the standard result

\be P_q (B, \mu=0,  T\to \infty)  = \frac{ 7 \pi^2 N_c T^4}{180}, ~~ \bar P_q =
\sum_q P_q =\frac{ 7 \pi^2 N_c T^4}{180}n_f.
 \label{13}\ee

 At this point one should stress the importance of the explicit summation over
 $n$, especially when $\mu\neq 0$. This indeed can be done as in \cite{15} with
 the result given in (\ref{7}).

 Finally we should comment on the accuracy of our representation (\ref{8}),
 (\ref{9}), which is obtained, when the summation over $n_\bot$ with $\sigma
 =-1$  in  (\ref{5}) is performed, using the Euler-Mc Laurent approximation (see
 \$ 59 of \cite{23}  for a discussion)
 \be \sum_{n_\bot =0}^\infty F(n_\bot +
 \frac12) \cong \int^\infty_0 F(x) dx + \frac{1}{24} F'(0),\label{14}
 \ee
  which yields the first term in the curly brackets in (\ref{8}), (\ref{9}).
  This substantiates  the good accuracy of the total expression  in the  whole
  region  of parameters except for a narrow region $T\ll m_q$, $ T\la
  \frac{{e_q B}}{2m_q} \ll \mu - m_q \equiv \mu_0$ where an oscillating regime
  sets in, considered in the next sections. As  an additional check of this
  accuracy we show in the next section that in the  expansion  of $P_q$ in powers
  of ($e_qB)^k$ the terms with $k=1 $ and  3 vanish identically.

  One finds that  (\ref{14}) is accurate within the terms $O\left( \frac{\sqrt{
  m^2_q+ e_qB}}{T}\right)$, when $\sqrt{ m^2_q + e_q B} \ll T$, while in the
  opposite case the sum (\ref{14}) is much smaller than the term with
  $\sigma=1$.

  \section{Magnetic susceptibility of the quark matter}

  In this section we are specifically interested in the $e_qB$ dependence of
  $P_q (B, \mu, T)$ and first of all in the quadratic term of this
  expansion --
  the magnetic susceptibility (m.s.). To this end we are exploiting the
  integral representation of $K_n$ \be  K_\nu (z) = \frac12 \left(\frac{2}{z}
  \right)^\nu\int^\infty_0 e^{-t - \frac{z^2}{4t}} t^{\nu-1}  dt, ~~K_\nu=K_{-\nu},
  \label{15}\ee
  which allows one to write expansions
  \be \frac{2T}{n} (e_q B + m^2_q) K_2 \left( \frac{n \sqrt{ e_q B+
  m^2_q}}{T}\right) = \frac{2T m^2_q}{n} \sum^\infty_{k=0} \left( \frac{e_q
  Bn}{2T m_q}\right)^k \frac{(-)^k}{k!} K_{k-2} \left(
  \frac{nm_q}{T}\right),\label{16}\ee
  \be K_0 \left( \frac{ n \sqrt{ m^2_q + e_q B}}{T} \right) = \sum^\infty_{k=0}
  \left( \frac{ ne_q B}{2Tm_q} \right)^k \frac{(-)^k}{k!} K_k \left(
  \frac{nm_q}{T}\right)\label{17}\ee
  and as a result Eqs. (\ref{8}), (\ref{9}) assume the form
  \be P_q (B, \mu, T)-P_q (0, \mu, T) = \frac{(e_qB)^2 N_c}{2\pi^2} \sum^\infty_{n=1} (-)^{n+1}
  \frac{(L_\mu^n + L^n_{-\mu})}{2} f_n,\label{18}\ee
  \be f_n = \sum^\infty_{k=0} \frac{(-)^k}{k!}\left( \frac{ ne_q B}{2Tm_q} \right)^k  K_k \left(
  \frac{nm_q}{T}\right)  \left[ \frac{1}{(k+1)(k+2)} - \frac16 \right].\label{19}\ee

  Note, that the linear term in (\ref{16}) exactly cancels the term $m_q K_1
  \left( \frac{nm_q}{T}\right)$ in (\ref{8}), so that the sum in (\ref{18})
  starts with the quadratic term, also the cubic term vanishes in (\ref{19}).
  Hence one can define the m.s. $\hat \chi_q$

  \be P_q (B, \mu, T) - P_q (0, \mu, T)= \frac{ \hat \chi_q}{2} (e_q B)^2 + O((e_qB)^4
  ).\label{20}\ee

  As the result, one arrives at the following expression for $\hat \chi_q$ \be
  \hat \chi_q( T ,\mu) = \frac{N_c}{3\pi^2} \sum^\infty_{n=1} (-)^{n+1}
  \frac{L_\mu^n + L^n_{-\mu}}{2} K_0 \left( \frac{nm_q}{T}\right),\label{21}\ee
  with $L_\mu \equiv L\exp \frac{\mu}{T}$.

  As the next step we are using for $K_0(z)$ the relation
  \be K_0 \left( \frac{nm_q}{T} \right) = \frac12 \int^\infty_0 \frac{dx}{x}
  e^{-n \left( \frac{1}{x} + \frac{m^2_q x}{4T^2}\right)}\label{22}\ee
  and obtain the final expression, summing over $n$,
  \be \hat \chi_q (T, \mu) = \frac{N_c}{3\pi^2}\frac{ J_q (\mu) + J_q
  (-\mu)}{2}, ~~ J_q(\mu) = \frac12 \int^\infty_0 \frac{dx}{x} \frac{L_\mu
  e^{-\left(\frac{1}{x} + \frac{m^2_q x}{4T^2}\right)}}{1+L_\mu e^{-\left(\frac{1}{x} + \frac{m^2_q
  x}{4T^2}\right)}}.\label{23}\ee

The total m.s. $\hat \chi (T, \mu)$ for a quark ensemble with
$n_f$ species is defined as \be \hat \chi(T, \mu) = \sum_q \hat
\chi_q (T,\mu) \left( \frac{e_q}{e}\right)^2.\label{24}\ee Note,
that one can define a more general form, appropriate for the
comparison with numerical simulations, when one simply extracts
the quadratic term $(e_qB)^2$, leaving m.f. nonzero in the rest
terms, namely \be \hat \chi_q (B, T,\mu) = \frac{2P_q (B,\mu,
T)}{(e_qB)^2} = \frac{ N_c}{3\pi^2}\sum^{\infty}_{n=1}
(-)^{n+1}\frac{L_\mu^n + L^n_{-\mu}}{2}\varphi_n (m^2_q +
e_qB)\label{25}\ee with \be \varphi_n (m^2_q + e_qB) =
\textrm{ln}\left(\frac{ 2T}{n\sqrt{e_q B+ m^2_q}}\right)+0.33.\label{26}\ee
One can see in (\ref{25}), (\ref{26}), that $m^2_q$ enters in
$\hat \chi_q (B,T,\mu)$ always in combination with $e_q B$, so
that one can define an effective mass \be (m^2_q)_{eff} = m^2_q+
e_qB,\label{27}\ee and $e_qB$ is of the order of the minimal m.f.
present in the lattice measurement of $\hat \chi_q$, which is
usually larger, than $m_u^2, m_d^2$.

Note, however, that the series over $n$ in  Eq. (\ref{25}) is not well
convergent  for $L_\mu > 1$  and one  should do in this case an explicit
summation over $n$ yielding (\ref{23}).

One can see in (\ref{21}), (\ref{25}), that at large $T\gg m_q,  \sqrt{m^2_q +
e_q B}$, each term in (\ref{21}), (\ref{25}) behaves as $\sim \ln
\frac{T}{m_q}$, implying that  $\chi_{q_1} >\chi_{q_2}$, when $m_{q_2}>
m_{q_1}$. However,  summation over $n$ yields in (\ref{23})   the denominator
which flattens the logarithmic grows  of the first term in the sum.  This
situation is especially interesting for the free case, when $L_\mu \equiv 1 \, (\mu=0)$,  in which case $ \hat
\chi_q^{(0)} \approx \frac{N_c}{3\pi^2}\sum_n (-1)^{n+1} K_0 \left(\frac{nm_q}{T}\right) $ and
summation over  $n$ leads to the  Eq.~(\ref{23}) with $L_\mu =1$. The
numerical result for the  first term $\hat \chi_q^{(0)}  (n=1) =
\frac{N_c}{3\pi^2} K_0 \left(\frac{m_q}{T}\right)$ and  for  the whole sum is shown in Fig.~1. The corresponding expression of  $\hat \chi^{(0)}_q $ for the  electron gas, which is twice  as small, can be found in \cite{19**},
and in \cite{27}.
\begin{figure}[h]
  \centering
  \includegraphics[width=9cm]{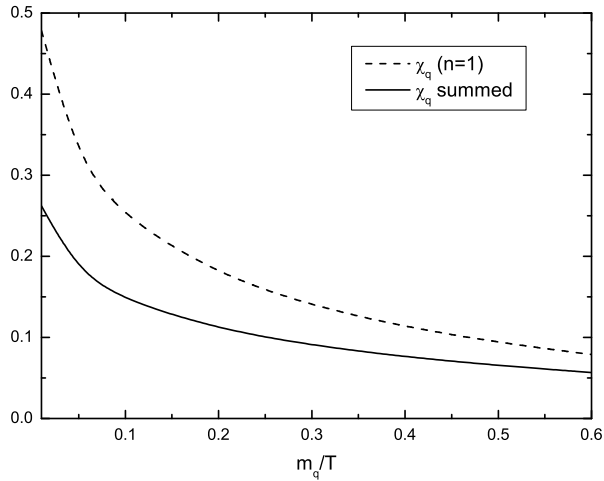}
  \caption{Magnetic susceptibility in SI units ($\chi_q = \frac{4\pi}{137}\hat\chi_q$) in free case, $L_\mu=1$, Eq.~(\ref{23}) (solid line), in comparison with the first term in the sum (\ref{21}).}
\end{figure}

As the next step one must define the Polyakov line, which in the neighborhood
of $T_c$ was found analytically in \cite{38,40} as \be L\equiv L^{(V)}(T) =
\exp \left( - \frac{V_1(\infty, T)}{2T}\right), ~ V_1 (\infty, T)\approx V_1
(\infty, T_c) =0.5 ~{\rm GeV}.\label{28}\ee
 Note, that by derivation in \cite{36} the Polyakov line $L^{(V)} (T)$ takes
 into account only the single quark interaction with the vacuum, given by
 $V_1(\infty, T)$, hence the superscript $V$, while on the lattice \cite{41}
 one measures the full Polyakov line, which can be expressed via the free
 energy $F_1 (\infty, T),$
 \be L^{(F)} (T) =\exp \left(- \frac{F_1 (\infty, T)}{2T}\right).\label{29}\ee

 As argued in \cite{40}, $F_1 < V_1$ and hence $L^{(F)}(T) > L^{(V)}(T)$. In
 \cite{18} both forms of $L(T)$ have been used for comparison with lattice data
 for m.s. without chemical potential.

It is interesting to compare $\hat \chi_q (T,\mu)$ for three different sorts of
quarks, $u,d,s$. Using (\ref{23}) with $m^2_q \to m^2_q (eff)$, one can find
three curves for $m_u (eff)=68$ MeV,  $m_d(eff) =49$ MeV,  $m_s(eff) = 111$ MeV and $\mu=0$,
which are in good agreement with the lattice data from \cite{19**}, see Fig.~2 (left graph), where we use $L=L^{(V)}$ from (\ref{28}). The sum of different quarks contributions is shown on Fig.~2, right graph.
\begin{figure}[h]
  \centering
  \includegraphics[width=6.7cm]{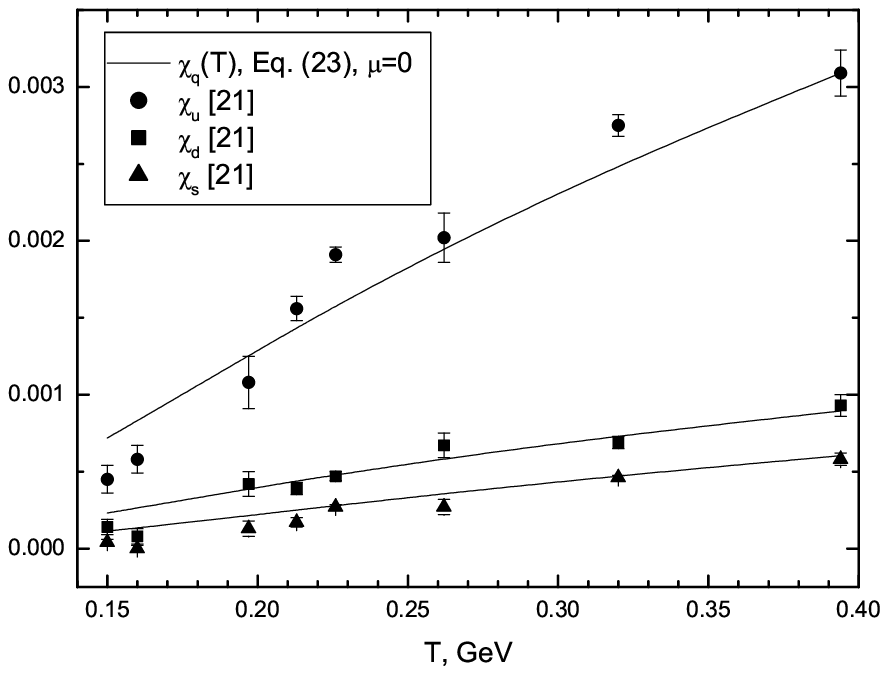}
    \includegraphics[width=6.7cm]{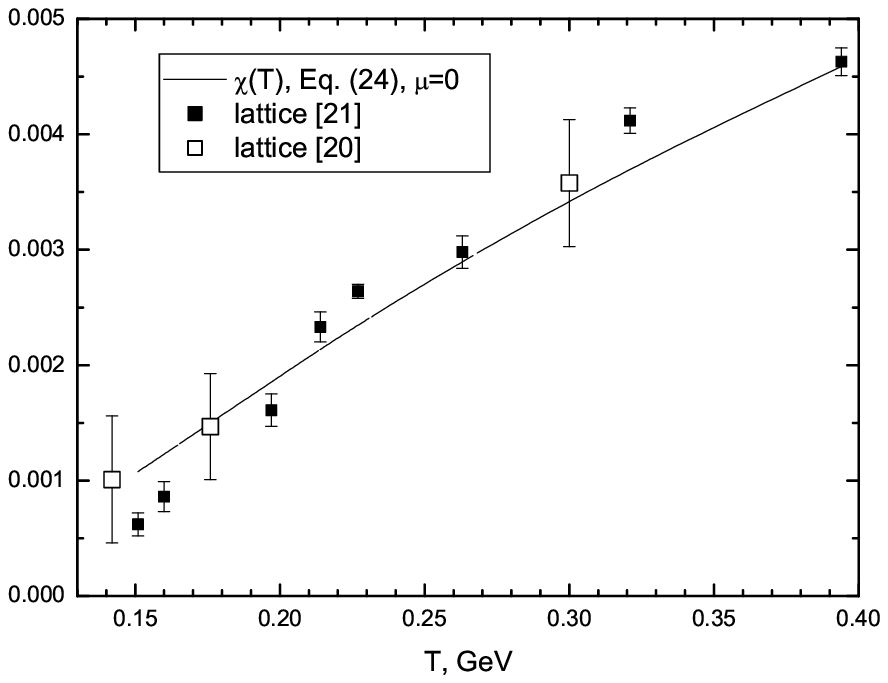}
  \caption{Magnetic susceptibility in SI units ($\chi_q = \frac{4\pi}{137}\hat\chi_q$) as a function of temperature for different sorts of quarks (left graph) and the total magnetic susceptibility (right graph) for the case $\mu=0$ in comparison with lattice data \cite{19*,19**}.}
\end{figure}

  \section{ The case of nonzero chemical potential}

  For $\mu>0$ one can use the standard representation (cf. Eq. ({24}) of
  \cite{15})
  \be P_q (B) = N_c T \frac{e_qB}{2\pi} (\psi (\mu) + \psi(-\mu)),\label{16g}\ee
  where \be \psi (\mu) = \sum_{n_\bot \sigma} \int^\infty_{-\infty}
  \frac{dp_z}{2\pi} \ln \left ( 1+ \exp \left( \frac{\bar \mu -
  E^\sigma_{n_\bot} (B)}{T}\right)\right),\label{17g}\ee
  and

  \be \bar \mu = \mu - \bar J = \mu - \frac{V_1(\infty, T)}{2}, ~~ \bar L_\mu
  = \exp \left( \frac{\bar \mu - \bar J}{T} \right)= \bar L \exp \left(\frac{\bar\mu}{T}\right),\label{18g}\ee
\be E^\sigma_{n_\bot} (B) = \sqrt{ p^2_z + ( 2n_\bot +1- \sigma)
e_q B + m^2_q}. \label{19g}\ee

Separately out in (\ref{17g})  the term $\sigma =1, n_\bot=0$ one can rewrite
$\psi(\mu)$ as (see Appendix of \cite{15} for details) \be \psi(\mu) =
\frac{1}{\pi T} \left\{ \phi (\mu) + \frac23 \frac{\lambda (\mu)}{e_q B} -
\frac{e_q B}{24} \tau (\mu) \right\},\label{20g}\ee
 where $\phi(\mu)$ does not depend on $e_q B$,
 \be \phi(\mu) = \int^\infty_0 \frac{p_z dp_z}{1+ e^{\frac{p_z-\bar \mu}{T}}},
 \label{21g}\ee

 \be \lambda(\mu) = \int^\infty_0 \frac{p^4 dp}{\sqrt{ p^2 + \tilde m^2_q}} \frac{1}{1+ \exp
 \left(\frac{{\sqrt{ p^2 + \tilde m^2_q}}-\bar \mu}{T}\right)},\label{22g}\ee

\be \tau(\mu) = \int^\infty_0 \frac{ dp }{\sqrt{ p^2  + \tilde m^2_q}\left(1+
\exp
 \left(\frac{{\sqrt{ p^2 + \tilde m^2_q}}-\bar
 \mu}{T}\right)\right)},\label{23g}\ee and $\tilde m^2_q = m^2_q + e_qB$.

It is clear, that with 4 dimensionful parameters $\bar \mu, m_q, T, e_q B$ one
has more than 6 limiting regions. Therefore in this section we shall confine
ourselves to only three situations, out of which two  were treated in \cite{23}
for the electron gas and called there  a) the case of weak fields, \be T\ll
\varepsilon_F = \mu_0, ~~ T\ll m_q,~~ \frac{eB}{2m_q}\ll T,\label{24g}\ee
 and b) the case of strong fields,
 \be T\la \frac{eB}{2m_q} \ll \mu_0, ~~ T\ll m_q, ~~ \mu_0.\label{25g}\ee
 In addition, there is another interesting region, namely $T\gg \mu, e_q B <
 T^2$, leading some access to the numerical simulations, which will be
 considered  now, while the  cases a) and b) are discussed in the  next  section.

We consider here the case of small $\mu, \mu\ll T$, and small m.f.,
$\sqrt{eB}\ll T$, when the possible region of oscillations due to the sum over
integrals $n_\bot$ in (\ref{2}) is unimportant, and one can replace the sum by
the integral, as it is done in (\ref{8}), (\ref{9}) using (\ref{14}). In this
case one can use (\ref{23}) with $L_\mu$ given by (\ref{9}) and $L(T)$ due to
(\ref{28}), (\ref{29}). We note here, that the influence of $\mu$ on $L(T)$ is
expected here to be negligible, see e.g. lattice data in \cite{42}.

In Fig.~3  we show a typical behavior of $\hat \chi_q (T,\mu)$, given by
(\ref{23}) for $L(T) =  L^{(V)} (T)$  from (\ref{28}). For $m_u = 68$ MeV (as for Fig.~2) and $\mu\equiv
\mu_u=(0,100,200)$ MeV one can see a set of curves $\hat \chi_u $ as a
function of $T$ in the interval (150-400) MeV.
\begin{figure}[h]
  \centering
  \includegraphics[width=9cm]{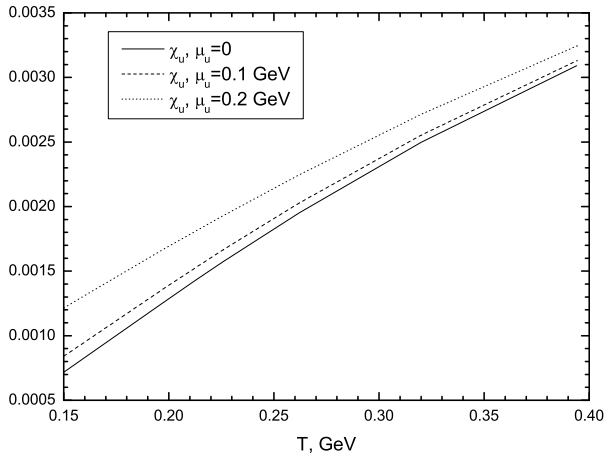}
  \caption{Magnetic susceptibility $\chi_u$ from (\ref{23}) in SI units ($\chi_q = \frac{4\pi}{137}\hat\chi_q$) as a function of temperature for nonzero values of chemical potential $\mu$.}
\end{figure}

Another possible characteristics of the small $\mu$ influence is the quark
number susceptibility of $\hat \chi_q (T, \mu_q)$, given by \be \hat
\chi_q^{(\mu)}(T) \equiv \left.\frac{\partial^2 \hat \chi_q (T,\mu)}{\partial
\mu^2} \right|_{\mu=0} = \frac{N_c T^2}{6\pi^2} \left(\left. \frac{\partial^2
J_q (\mu)}{\partial \mu^2} + \frac{\partial^2 J_q
(-\mu)}{\partial \mu^2}\right)\right|_{\mu=0}.\label{30}\ee

Differentiating (\ref{23}) one obtains

\be \chi_q^{(\mu)} (T) = \frac{N_c}{12\pi^2  } \int^\infty_0 \frac{dx}{x}
\frac{L_\mu
  e^{-\left(\frac{1}{x} + \frac{m^2_q x}{4T^2}\right)}
   \left( 1- L_\mu e^{-\left(\frac{ 1}{x} +  \frac{m^2_q x}{4 T^2}\right)}\right)}{ \left(1+L_\mu e^{-\left(\frac{1}{x} + \frac{m^2_q
  x}{4T^2}\right)}\right)^3} + (\mu\to - \mu).\label{31}\ee

This combined quark-number and magnetic susceptibility is a generalization of
the powerful technic of the study of the chemical potential influence on
thermodynamic potentials and phase transition on the  lattice, (see a recent
paper \cite{43} for a discussion and references).

One should note, that the corresponding quark number susceptibility (q.n.s.)
was calculated for zero m.f. in the framework of our approach in \cite{37}. To
this end one can use (\ref{16g}), (\ref{22g}),  since only $\lambda(\mu)$
survives for   $e_q B\ll m_q$, and one has \be P_q (B=0) =
\frac{N_c}{3\pi^2}\left\{ \lambda (0)+ \sum^\infty_{k=2} \frac{1}{k!}
\frac{\partial^k \lambda(\mu-q)}{\partial \left( \frac{\mu_q}{T}\right)^k}
\left( \frac{\mu_q}{T} \right)^k\right\}\label{41*}\ee and $\lambda(\mu_q)$  is
given by (\ref{22g}).

The equivalent series for $\hat\chi_q (T,\mu)$ is obtained by the replacement
in (\ref{41*}), $\lambda(\mu_q) \to \frac{1}{2} (J_q(\mu) + J_q (-\mu))$.

\section{The cases of strong and weak fields}

We consider now the cases of the weak and strong fields, Eqs. (\ref{24g}) and
(\ref{25g}) respectively, essentially the material of \S\S~58,59 of \cite{23}.

 Note, that our $\mu$ for quarks contains the quark mass, $\mu=m_q+\mu_0$,
 while $\mu_0$ depends on density, $\mu_0=\varepsilon_F$ for electron gas.

 We start with the case a) $ \frac{eB}{2m_q} \ll T \ll m_q, \mu_0$.

 Here one can use (\ref{16g}), (\ref{20g}) and take into account, that the
 quadratic in $e_q B$ terms come only from $\lambda(\mu)$ and $\tau(\mu)$. We
 neglect the terms $O\left( \frac{e_qB}{m^2_q}\right)  $ (and hence $\psi(-\mu)$ in (\ref{16g}))
 and  omitting $V_1(\infty, T)$ write the exponent in the integrand of (\ref{22g}) as  $$\exp
 \left(\frac{{\sqrt{ p^2 + \tilde m^2_q}}-\bar
 \mu}{T}\right) = \exp \left( \frac{p^2}{2m_q T} - \frac{\bar
 \mu_0}{T}\right), \quad \bar \mu = \mu_0 + m_q,
 ~~$$ \be\bar \mu_0 = \mu_0  - \frac{e_q B}{2 m_q}\equiv
 \mu- m_q  - \frac{e_q B}{2m_q},\label{26g}\ee
and the integral (\ref{22g}) can be rewritten, using the variable $z=
\frac{p^2}{2m_qT}$, \be \lambda(\mu) = T^4 \left( \frac{2m_q}{T} \right)^{3/2}
\int^\infty_0 \frac{z^{3/2} dz}{e^{z-\bar \mu_0/T} +1}.\label{27g}\ee The
integral on the r.h.s. of (\ref{27g}) is exactly of the type considered in
\cite{23}, \S~58, where the asymptotic series was obtained in powers
$\left(\frac{\bar \mu_0}{T}\right)^k$. Keeping the leading term, one has \be
\lambda(\mu) = T^4 \left( \frac{2m_q}{T} \right)^{3/2} \left( \frac25
\left(\frac{ \bar \mu_0}{T} \right)^{5/2} + O\left(\frac{\bar \mu_0}{T}
\right)^{1/2} \right).\label{28g}\ee

Expanding $\bar \mu_0=\tilde \mu_0 -\frac{e_qB}{2m_q}$, and
keeping the term $\left(\frac{e_qB}{2m_q}\right)^2$, one obtains
from (\ref{16g}), (\ref{20g}) the paramagnetic contribution to
$P_q$

\be P_q^{(2)} = \frac{ N_c (e_q B)^2}{4\pi^2}
\left(\frac{\mu_0}{2m_q} \right)^{1/2}, \label{29g}\ee and for
$\tau(\mu)$ one has \be \tau(\mu) = \sqrt{\frac{m_qT}{2}} 2 \left(
\frac{\bar \mu_0}{T}\right)^{1/2}.\label{a29g}\ee

Inserting these values  into (\ref{16g}), (\ref{20g}) one obtains for  the
$\tau(\mu)$  the contribution to  $\hat \chi$, which is $\left(
-\frac13\right)$ of the contribution of $\lambda(\mu)$ \be \chi =
\frac{N_c}{3\pi^2} \left( {e_q}\right)^2 \sqrt{\frac{\mu_0}{2m_q}},
\label{30g}\ee which coincides with the total m.s. of the electron gas in the
weak m.f., given in \cite{23}, when $m_q=m_e, N_c=1, \mu_0 = \varepsilon_F$.

The same result can obtained directly from (\ref{23}) inserting there $L_\mu =
\exp \left( \frac{\mu_0 + m_q}{T} \right)$ , and using instead of $x$ the
variable $\frac{\varepsilon}{T} = \frac{1}{x} + \frac{ m^2_q}{4 T^2} x -
\frac{m_q}{T}$. In the limit $T\ll m_q$ one obtains $J_q \approx
\sqrt{\frac{2}{m_q}} \int^\infty_0 \frac{d\varepsilon}{\sqrt{\varepsilon}}
\frac{1}{1+e\frac{\varepsilon -\mu_0}{T}}$, which using the same technic as in
(\ref{28g}) yields $\sqrt{\frac{2\mu_0}{m_q}}$  and one gets the result
(\ref{30g}).

 We now turn to the  case b), $ T\la \frac{e_q B}{2m_q} \ll \mu_0$, which is
interesting for us, since it provides the oscillating behavior, which is not
present in our form (\ref{17g}), see discussion in \S~59 of \cite{23}.

Indeed, the form (\ref{20g}) obtains, when one considers $e_qB$ outside of the
interval  b), or else,  when one averages the result over some interval of
$e_qB$, comprising many values of $n_\bot$ in (\ref{17g}).

To this end we rewrite $\psi(\mu)$ in (\ref{17g}), separating the first term
$\sigma =-1$, not depending on m.f. and use the Poisson formula \be \frac12
F(0) + \sum^\infty_{n=1} F(n) = \int^\infty_0 F(x) dx + 2 Re \sum^\infty_{k=1}
\int^\infty_0 dx F (x) e^{2\pi i kx}, \label{31g}\ee where \be  F(x) =
\int^\infty_{-\infty} \frac{dp_z}{2\pi}\ln \left(  1+ \exp \left( \frac{ \bar
\mu - \sqrt{ p^2_z+ m^2_q + 2 e_q Bx}}{T}\right)\right).\label{32g}\ee and
$\bar \mu  = m_q +\mu_0 - \frac{V_1(\infty, T)}{2}$ for quarks and $\bar \mu =
m_e + \mu_0$ for the electron gas.

 For small $T$, $ \frac{e_qB}{m_q}$ as
compared to $m_q$ (nonrelativistic situation), we write the exponent as
$\left(\mu_0 - \frac{p^2_z}{2m_q} - \frac{e_qBx}{m_q}\right)\frac{1}{T}$ and we
are in the exact correspondence with the equations in \S~60 of \cite{23}, when
one replaces our $\mu_0, m_q, e_q$ by $\mu,m,e$ of the electron gas. The
resulting expression for $P_q (B)$ (\ref{16g}) is \be P_q (B) = N_c
\frac{e_qB}{2\pi^2} (\phi(\mu) + \phi(-\mu)) - \frac{ N_c T (e_q
B)^{3/2}}{4\pi^2} \sum^\infty_{k=1} \frac{\cos \left( \frac{2\pi \mu_0 m_q
k}{e_q B} - \frac{\pi}{4}\right)}{k^{3/2} \textbf{sh} \left( \frac{2\pi^2 kT
m_q}{e_qB}\right)}.\label{33g}\ee

 One can expect for the quark gas the same oscillations as in the de Haas-van
 Alphen effect, but the period of oscillations for quarks in $e_q B$ is $\mu_0
 m_q$ and we assume $T\ll m_q$, hence this is improbable for a deconfined quark
 gas, where $T_c > m_q, ~q=u,d,s.$ Therefore  we shall try to proceed, assuming only that $    T\la\frac{e_q
 B}{2m_q} $ , but allowing for $T\gg m_q$. Then the exponent in (\ref{32g}) can
 be rewritten as \be \frac{\bar \mu^2 - m^2_q - p^2_z - 2e_q Bx}{T(\bar\mu +
 \sqrt{m^2_q + p^2_z + 2  e_q Bx})}\equiv \frac{\delta \mu^2}{T} -\frac{(p^2_z+
 2 e_q Bx)}{2M_qT},\label{51g}\ee where $\delta\mu = \frac{\bar \mu^2 -
 m^2_q}{2 M_q}$ and $2 M_q  \equiv  \bar \mu + \sqrt{m^2_q +p^2_z +2 e_q Bx}$.

Approximating $M_q $ by some average value,  not depending on $p_z$, one can
 exploit the  final  Eq. (\ref{33g}), replacing there $m_q$ by $M_q$ and  $M_0$
 by $\delta\mu$. As a result one expects the  oscillations of $P_q (B)$ for
 growing $e_q B$ for $\frac{e_q B}{M_q T} \ga 1$ and $T\ll \delta\mu$.

Indeed, the   oscillating term in the integral  (\ref{17g}), using (\ref{31g}),
(\ref{32g}) can be written as (cf. \S~60 of \cite{23})

\be I_k =- e_q B \int^\infty_{-\infty} \int^\infty_0 \ln \left[ 1+\exp \left(
\frac{\delta \mu}{T} -\frac{p^2_z+ 2e_qBx}{2M_qT}\right)\right] e^{2\pi i kx} d
p_z dx.\label{52g}\ee

Introducing new variable $\varepsilon = \frac{p^2_z +2 e_q Bx}{2M_q} $ instead
of $x$, one obtains for the oscillating part of (\ref{52g}) \be \bar I_k
=-\int^\infty_{-\infty} \int^\infty_0 \ln \left[ 1+\exp \left( \frac{\delta
\mu-\varepsilon }{T} \right) \right] \exp \left(\frac{ 2 i \pi k \varepsilon
M_q}{e_q B} \right) \exp \left(-\frac{ i\pi k p^2_z}{e_qB}\right) d\varepsilon
dp_z.\label{53g}\ee

In (\ref{53g}) the essential part of integration region is $p^2_z \sim e_q B$,
while for the oscillating regime  $\delta \mu \sim \varepsilon, \delta \mu \gg
\frac{e_q B}{2M_q}$, therefore one replaces the lower limit of the
$\varepsilon$ integration by zero. Moreover, $2M_q \approx \bar \mu + \sqrt{
m^2_q + 2e_q Bx} \ga \bar \mu \gg m_q$, and $\delta\mu \approx \bar \mu$.

Hence the final form of the oscillating part of the thermodynamic potential can
be written instead of (\ref{33g}) as \be \Delta P_q (B) =- N_c  \frac{ (e_q
B)^{3/2}T}{4\pi^2}  \sum^\infty_{k=1} \frac{\cos \left( \frac{2\bar \mu^2}{e_q
B} k - \frac{\pi}{4}\right)}{k^{3/2}  \textrm{sh} \left(\frac{\pi^2 k T \bar \mu}{e_q B}
\right)}.\label{54g}\ee

One can see, that the relativistic quark gas potential (\ref{54g}) contains a
much larger denominator  due to $T\gg m_q$ , as  compared to the electron gas
potential (\ref{33g}), leading to a relatively smaller  amplitude of
oscillations.

\section{Summary and perspectives}

We have developed above the theory of m.s. of the quark-antiquark matter in
m.f., based on the explicit expressions for  the thermodynamic potentials
obtained by us in \cite{15}. The case of m.s. for zero chemical  potential
$\mu$ was studied in our previous paper \cite{18}, where it was shown, that
m.s. $\hat \chi (T)$ is a strong function of $T$, growing with $T$ due to
Polyakov line factors. This behavior agrees well with recent lattice
calculations \cite{19,19*,20}, when one takes into account a possible
modification of the effective quark mass as in (\ref{27}). In the present paper
we further examined the zero $\mu$ m.s., calculating m.s. for  different quarks
$(u,d,s)$ and  comparing with lattice data of \cite{19**} in our  Fig.~2. As an
additional topic we consider the free quark-antiquark gas m.s., which obtains
from (\ref{23}) putting $L_\mu \equiv 1$, and compare it with the corresponding
m.s. of the  electron gas from \cite{28}. We observe a strong modification  of
the result due to the sum over Matsubara numbers. The main part of our results
belong to the case of nonzero chemical potential $\mu$ in sections 4 and 5.
Here m.s. $\hat \chi (T, \mu)$ has different behavior in the regions of small
and  large $\mu$, $\mu\ll T$ and $\mu\gg T$. In the  first case, considered in
section 4, one can define the double magnetic- quark number susceptibilities.
In the case of large $\mu, \mu\gg T$ and $\frac{eB}{2m_q}\ll T$, one obtains
the standard Pauli paramagnetism  \cite{21} and Landau diamagnetism \cite{22}
contributions to the m.s. given in (\ref{a29g}).

Finally, in the case of large $\mu$ and large m.f. one arrives at the Landau
theory of the de Haas-Van Alphen effect, written for nonrelativistic quarks in
(\ref{32g}). The generalization to the case of relativistic quarks for $T\gg
m_q$ is obtained in (\ref{53g})  and shows much milder amplitude of
oscillations with growing $e_qB$. As it is we have developed the full theory of
m.s. of the quark gas interacting  with the QCD vacuum in the so-called Single
Line Approximation  (SLA) \cite{36}, when the interaction enters in the  form
of Polyakov lines. This allows to obtain m.s. at zero or small $\mu$, and a
good agreement
 was found with lattice data at least in the first case.  In this approximation
 the interquark interactions are disregarded, however at  larger $\mu$ (and
 hence larger quark densities) this effect can become important and this was
 discussed in \cite{36,40}. In SLA the QCD phase diagram in the  $\mu-T$ plane
 was found in \cite{15} and does not contain critical points. However for
 larger $\mu$ the interquark interaction becomes important and depends on $\mu$
 both in the confined \cite{43a} and  deconfined \cite{38,44} states. As a
 result the problem of the quark-hadron (qh) matter transitions should be
 solved with the full account of the  interquark  (beyond SLA) interactions.
 One aspect of this transition -- the  formation of the multiquark states and
 the nucleon matter was considered in \cite{45}, and shown to be important for
 the quark cores of neutron stars.

 The authors are grateful to   M.~D'Elia and G.~Endrodi for stimulating correspondence.

The RFBR grant  1402-00395 is gratefully acknowledged.

\end{document}